\documentclass[letterpaper,12pt,leqno]{article}
\usepackage{paper,math}
\pdfoutput=1
\def\bib{paper.bib}
\def\pdf{figures.pdf}
\published{Oxford Economic Papers}{https://www.pascalmichaillat.org/7.html}
\hypersetup{pdftitle={An Economical Business-Cycle Model}}

\begin{document}

\title{An Economical Business-Cycle Model}
\author{Pascal Michaillat, Emmanuel Saez
\thanks{Michaillat: Brown University. Saez: University of California--Berkeley. This work was supported by the Economic and Social Research Council [ES/K008641/1]; the Institute for New Economic Thinking; the British Academy; and the Berkeley Center for Equitable Growth. We thank Regis Barnichon, Giancarlo Corsetti, Wouter den Haan, Emmanuel Farhi, Jordi Gali, Yuriy Gorodnichenko, Pawel Kopiec, Etienne Lehmann, Adam McCloskey, Karel Mertens, Yoshiyasu Ono, Kevin Sheedy, Carl Walsh, Johannes Wieland, Michael Woodford, and Francesco Zanetti for helpful discussions and comments.}}
\date{April 2021}

\begin{titlepage}\maketitle

This paper develops a new model of business cycles. The model is economical in that it is solved with an aggregate demand-aggregate supply diagram, and the effects of shocks and policies are obtained by comparative statics. The model builds on two unconventional assumptions. First, producers and consumers meet through a matching function. Thus, the model features unemployment, which fluctuates in response to aggregate demand and supply shocks. Second, wealth enters the utility function, so the model allows for permanent zero-lower-bound episodes. In the model, the optimal monetary policy is to set the interest rate at the level that eliminates the unemployment gap. This optimal interest rate is computed from the prevailing unemployment gap and monetary multiplier (the effect of the nominal interest rate on the unemployment rate). If the unemployment gap is exceedingly large, monetary policy cannot eliminate it before reaching the zero lower bound, but a wealth tax can. (JEL: E19, E24, E32, E43, E52, E62, E71)

\end{titlepage}
\section{Introduction}

\paragraph{Limitations of the New Keynesian model} The New Keynesian model is currently the canonical model of business cycles \cp{G15}. Yet, it arguably lacks two attributes of a good scientific model \cp[pp.~36--41]{K57}.

First, the New Keynesian model is not economical. It is not the model that leading macroeconomists have committed to memory and use for day-to-day thinking about macroeconomic issues. Instead, they turn to the old IS-LM model \cp{K00,Kr18}. It is not the model taught to undergraduates either. Instead, popular intermediate-macro textbooks continue to explain business cycles using the IS-LM model \cp{ABC17,M19}. Last, perhaps because of its unwieldiness, the model has not been used outside macroeconomics by those in related fields.

Second, the New Keynesian model does not describe business cycles sufficiently well. It does not feature unemployment, although high unemployment is the principal problem caused by recessions \cp{BG08}. And it makes anomalous predictions about long-lasting zero-lower-bound (ZLB) episodes \cp{MS18}. These anomalies have become more salient as these episodes have become more common.

\paragraph{Attributes of this paper's model} This paper develops a model of business cycles that is both economical and descriptive. The model is economical in that it is solved with an aggregate demand-aggregate supply diagram, and the effects of shocks and policies are derived by comparative statics. The most complicated step of the analysis is the derivation of the consumption Euler equation. The model is descriptive in that it features unemployment, which fluctuates in response to business-cycle shocks, and it behaves well during long-lasting or permanent ZLB episodes. 

\paragraph{Relation to previous research} To build the model, we combine two unconventional assumptions. 

To generate unemployment, we assume that producers and consumers meet through a matching function. We borrow this assumption from \ct{MS13,MS15}. The models in \ct{MS13,MS15} do not feature interest-bearing assets, however, so they cannot be used to think about monetary policy or the ZLB.

To accommodate permanent ZLB episodes, we assume that relative wealth enters the utility function. We borrow this behavioural assumption from \ct{MS18}. The model in \ct{MS18} is New Keynesian, however, so it does not feature unemployment.

\paragraph{Service economy} The model features households who sell labour services, spend their income on services sold by other households, and save unspent income using government bonds. The focus on services allows us to merge the labour and product markets into a single market and thus simplify the analysis. Such focus also seems realistic. In the United States for example, the share of employment in service-providing industries is $80.3\%$ in 2019, and it is projected to increase further in the future.\footnote{Bureau of Labor Statistics,`Employment by Major Industry Sector,' last modified September 1, 2020, \url{https://perma.cc/3T8A-R8L5}.}

\paragraph{Matching function} The first key assumption of the model is that producers and consumers of services meet through a matching function. Because of the matching function, the model features unemployment. Every day, people are available to work for a certain number of hours and produce a certain amount of services. But they cannot always find customers to hire them, so they are not always working.

Although in standard models markets are perfectly or monopolistically competitive, in reality most markets appear to be organized around a matching function. This is of course true of labour markets \cp{PP01}. But it is also true of many product markets \cp[pp.~519--521]{MS13}. For instance, in an average day in the United States, more than 15\% of firms' capacity remains idle. Moreover, the rate of idleness is sharply countercyclical, just like the rate of unemployment.

\paragraph{Wealth in the utility function} The second key assumption of the model is that households derive utility not only from consumption but also from their relative wealth. With wealth in the utility function, the Euler equation is modified because people save not only to smooth consumption over time but also to accumulate wealth. As a result, the aggregate demand curve is nondegenerate, which in particular allows for permanent ZLB episodes.

Although in standard models people only hold wealth to smooth consumption, in reality people appear to enjoy wealth over and above the future consumption it can provide \cp[section~2]{MS18}. For example, neuroscientists have found that wealth itself provides utility, independently of the consumption it can buy \cp{CLP05}. And economists, psychologists, and sociologists have documented that by accumulating wealth, people achieve high social status, which they relish \cp{WF98,F10,HF11,CT13,R14,AHH15,MKC17}. The wealth-in-the-utility-function assumption also explains why people save at single-digit interest rates while they exhibit double-digit discount rates \cp[section~6]{MS18}.

\paragraph{Price norm} Any model with a matching function needs to assume a price mechanism. The price mechanism determines how unemployment respond to shocks. If the price is flexible, the price responds to shocks but unemployment does not. If the price is rigid, on the other hand, the price does not respond to shocks but unemployment does.

In the United States, a lot of labour is idle in slumps and very little labour is idle in booms \cp[figure 2]{MS13}. In 2009, during the Great Recession, the rate of unemployment reached 10\%, and the rate of idleness reached 33\% in manufacturing industries and 20\% in non-manufacturing industries. In contrast, in 2000, at the peak of the dot-com bubble, the rate of unemployment was only 4\%, and the rate of idleness in all industries was only 13\%.

The only way to generate such fluctuations is to assume some price rigidity \cp{MS13}. Here, we simply assume that the price of services grows at a fixed rate. Following \ct{H05}, we interpret this fixed inflation rate as a social norm---which in the long run could be determined by communication from the central bank. 

A a fixed inflation rate is not unrealistic. In the United States, inflation responds neither to fluctuations in unemployment (\inp[figure 1]{SW10}; \inp[figure 1]{SW19}), nor to monetary-policy shocks \cp{CEE99}.

\paragraph{Solution of the model} The model is easy to solve. Output and market tightness are found at the intersection of an aggregate demand (AD) curve and an aggregate supply (AS) curve. The AD curve describes the trade-off between consuming and holding wealth given by the Euler equation. The AS curve describes the amount of output produced when unemployment is on the Beveridge curve. All the other variables are computed from output and tightness.

\paragraph{Business cycles} The AD-AS diagram also gives the response of the model to various aggregate demand and supply shocks: shocks to thriftiness, to labour productivity, or to labour-force participation. Negative aggregate demand shocks lead to lower output, lower tightness, and higher unemployment rate. Negative aggregate supply shocks lead to lower output, but higher tightness and lower unemployment rate. 

\paragraph{Inefficiency} The efficient unemployment rate is independent of aggregate demand and supply shocks, whereas the actual unemployment rate responds to these shocks. Unemployment fluctuations are therefore inefficient. When unemployment is inefficiently high, the amount of services produced and consumed is too low. When unemployment is inefficiently low, too many man-hours are devoted to recruiting, so the amount of services consumed is also too low. Despite such aggregate inefficiency, individual relationships are always bilaterally efficient, unlike in other rigid-price models \cp{B77,HR20}.

\paragraph{Monetary policy} The government conducts monetary policy by setting the nominal interest rate on bonds. Since it is optimal for monetary policy to eliminate the unemployment gap, the optimal nominal interest rate is given by a simple formula involving the prevailing unemployment gap and monetary multiplier (the effect of the nominal interest rate on the unemployment rate). The formula can be implemented because we have estimates of the unemployment gap and monetary multiplier \cp{CEE99,R15,MS16}. Applied to the United States, the formula indicates that the Federal Reserve should reduce the federal funds rate by $2$ percentage points for each percentage point of unemployment. This prescription is close to what the Federal Reserve does in practice \cp{BB92,SW01}.

\paragraph{Wealth tax at the ZLB} If the unemployment gap is excessively large, monetary policy cannot eliminate it before reaching the ZLB. At this point, the government can use a wealth tax because the wealth tax perfectly substitutes for monetary policy but is not subject to the ZLB. An adequate increase in the wealth tax rate eliminates the unemployment gap even at the ZLB.

\section{The Model}\label{s:model}

We begin by presenting our model. The model features unemployment, and it accommodates permanent ZLB episodes. The solution of the model is found at the intersection of a downward-sloping AD curve and an upward-sloping AS curve in a tightness-output plane. The AD curve derives from the Euler equation, and the AS curve from the Beveridge curve.

\subsection{Structure of the model}

The model is set in continuous time. It features households and a government. Households trade labour services and government bonds.

\paragraph{Households} There is a measure 1 of households, which are initially identical. Households are large, and we abstract from the randomness introduced by the matching process at the household level. As a result, we can analyse the model using a representative household.

\paragraph{Government} The government issues nominal bonds, sets taxes, and sets the nominal interest rate through the central bank.

\paragraph{Labour services} The only goods produced in the economy are labour services. There are no firms: households produce the services themselves and sell them to other households. On the service market, all trades are mediated by a matching function, and search is random.

\paragraph{Government bonds} To obtain an interesting concept of aggregate demand, households must have the choice between spending their income on services or on something else. Here households choose between buying services and buying government bonds.

\subsection{Supply of services}

We model the supply of services as in \ct{MS15}.

\paragraph{Labour force} The size of the labour force is $l>0$.

\paragraph{Labour productivity} Each worker in the labour force has the capacity to produce $a>0$ services per unit time. The parameter $a$ describes labour productivity.  

\paragraph{Employment relationships} Services are sold through long-term worker-household relationships. Once a worker has matched with a household, she becomes a full-time employee of the household. She remains so until they separate, which occurs at rate $\l>0$. As an employee, the worker produces $a$ services per unit time for the household. At time $t$, the services are sold at a unit price $p(t)$, so the worker's income is $a p(t)$. The inflation rate is $\pi(t)= \dot{p}(t)/p(t)$.

\paragraph{Unemployment} Because of the matching function, not all jobseekers find a job, so there is always some unemployment. The unemployment rate $u(t)$ is the share of workers in the labour force who are not employed by any households. Hence, the number of employed workers is 
\begin{equation}
n(t) = [1-u(t)] l,
\label{e:n}\end{equation}
and the aggregate output of services is
\begin{equation}
y(t) = a n(t) =  [1- u(t)] a l.
\label{e:y}\end{equation}
The aggregate productive capacity is $a l$: it is the amount of services that would be produced if all workers in the labour force were employed. Output is less than the capacity because some workers are unemployed.

\paragraph{Matching function} To find new employees, households advertise $v(t)$ vacancies. Based on the numbers of vacancies and jobseekers, a Cobb-Douglas matching function determines the number of new employment relationships formed per unit time: 
\begin{equation*}
m(t)=\m v(t)^{1-\h} [l-n(t)]^{\h},
\end{equation*} 
where $\m>0$ is the matching efficacy and $\h\in(0,1)$ is the matching elasticity.

\paragraph{Matching rates} With constant returns to scale in matching, the rates at which workers and households form new relationships is determined by the market tightness $\t(t)$. The market tightness is the ratio of the matching function's two arguments: 
\begin{equation}
\t(t) = \frac{v(t)}{l-n(t)}. 
\label{e:theta}\end{equation}
Each of the $l-n(t)$ unemployed workers finds a job at a rate
\begin{equation}
f(\t(t)) = \frac{m(t)}{l-n(t)}= \m \t(t)^{1-\h}, 
\label{e:f}\end{equation}
and each of the $v(t)$ vacancies is filled at a rate
\begin{equation}
q(\t(t))= \frac{m(t)}{v(t)}= \m  \t(t)^{-\h}. 
\label{e:q}\end{equation}
The job-finding rate $f(\t)$ is increasing in $\t$, and the vacancy-filling rate $q(\t)$ is decreasing in $\t$. Hence, when tightness is higher, it is easier to find a job and sell services, but harder to find a worker and buy services.

\paragraph{Unemployment dynamics} Given the matching process, the number of employment relationships evolves according to the following differential equation:
\begin{equation*}
\dot{n}(t) = f(\t(t)) \bs{l-n(t)} - \l  n(t),
\end{equation*}
where $f(\t(t)) \bs{l-n(t)}$ is the number of new relationships created at time~$t$, and $\l n(t)$ is the number of existing relationships dissolved at time~$t$. From equation~\eqref{e:n}, we infer that the unemployment rate also follows a differential equation:
\begin{equation}
\dot{u}(t) = \l  \bs{1-u(t)} - f(\t(t)) u(t).
\label{e:udot}\end{equation}

\paragraph{Beveridge curve}  The critical point of equation~\eqref{e:udot} is
\begin{equation}
u = \frac{\l}{\l+f(\t)}.
\label{e:u}\end{equation}
This negative relationship between unemployment rate and tightness is the Beveridge curve. It is the locus of unemployment and tightness such that the number of new employment relationships created at any point in time, $f(\t) u l$, equals the number of relationships dissolved at any point in time, $\l (1-u)l$.

The unemployment rates given by equations~\eqref{e:udot} and~\eqref{e:u} are indistinguishable (\inp[pp.~398--399]{Ha05}; \inp[p.~236]{Pi09}). In fact, \ct[p.~31]{MS16} find that when $\l$ and $f(\t)$ are calibrated to US data, the deviation between the two unemployment rates decays at an exponential rate of 62\% per month. This means that about 50\% of the deviation evaporates within a month, and about 90\% within a quarter. We therefore assume that the unemployment rate is given by equation~\eqref{e:u} at all times.

\subsection{Demand for services}

To model the demand for services, we combine elements from \ct{MS13}, who derive the demand for services in a static matching model, and elements from \ct{MS18}, who analyse consumption and saving when relative wealth enters the utility function.

\paragraph{Recruiting cost} To fill their vacancies, households allocate some of their employees to recruiting. Each vacancy requires $\k>0$ workers per unit of time. These workers provide recruiting services, such as advertising the vacancies, reading applications, and interviewing suitable candidates.

\paragraph{Consumption versus output} Recruiting services do not directly provide utility to households. We therefore distinguish between the amount of services that households purchase, $y(t)$, and the amount of services that provide utility, $c(t)< y(t)$. We refer to $c(t)$ as consumption; it is computed by subtracting recruiting services from output $y(t)$. 

\paragraph{Vacancies} When $v$ vacancies are posted, $q(\t) v$ new employment relationships are created at any point in time. On the Beveridge curve, that number equals the number of relationships that separate at any point in time, $\l n$. Hence, sustaining employment $n$ requires $v = \l n/q(\t)$ vacancies. The vacancies are managed by $\k \l n/q(\t)$ recruiters, who produce $a \k \l n/q(\t) = \k \l y/q(\t)$ recruiting services. 

\paragraph{Recruiting wedge} When $y$ services are produced, only $c = y  - \k \l y /q(\t)$ services are actually consumed. Consumption and output are therefore related by
\begin{equation}
y = \bs{1+\tau(\t)} c,
\label{e:c}\end{equation}
where 
\begin{equation}
\tau(\t) = \frac{\k  \l}{q(\t)-\k  \l}
\label{e:tau}\end{equation}
is the wedge between consumption and output caused by recruiting. The recruiting wedge $\tau(\t)$ is positive and increasing on $[0,\t_{\tau})$, where $\t_{\tau}$ is defined by $q(\t_{\tau}) = \k \l$; furthermore, $\tau(0)=0$ and $\lim_{\t\to \t_{\tau}} \tau(\t) = + \infty$. This implies that when tightness is higher, a larger share of services are devoted to recruiting.

\paragraph{Budget constraint} Without randomness at the household level, the representative household exactly sells a share $1-u(t)$ of its productive capacity $a l$; the remaining capacity, $u(t) a l$, is idle. From the sale of services, it earns $p(t) \bs{1-u(t)} a l$ at time $t$. 

The household also holds a nominal stock $b(t)$ of government bonds, which pay a nominal interest rate $i(t)$. The household earns $i(t) b(t)$ from the bonds at time $t$.

The household spends part of its income on services produced by other households. To consume 1 service, it purchases exactly $1+\tau(\t(t))$ services; the extra $\tau(\t(t))$ services are used for recruiting. Hence, to consume $c(t)$ services, it purchases $\bs{1+\tau(\t(t))} c(t)$ services, which cost $p(t) \bs{1+\tau(\t(t))} c(t)$.

Combining the household's income and spending, we obtain the flow budget constraint:
\begin{equation}
\dot{b}(t)= i(t) b(t) + p(t) \bs{1-u(t)} al  - p(t) \bs{1+\tau(\t(t))}  c(t) - T(t),
\label{e:dotb}\end{equation}
where $T(t)$ is a lump-sum tax (or transfer) that the government uses to balance its budget.

For convenience, we rewrite the flow budget constraint in real terms. We denote the household's real stock of bonds by 
\begin{equation}
w(t) = \frac{b(t)}{p(t)}
\label{e:w}\end{equation}
and the real interest rate by 
\begin{equation*}
r(t) = i(t)-\pi(t).
\end{equation*}
The growth rates of the real and nominal stocks of bonds are related by 
\begin{equation*}
\frac{\dot{w}(t)}{w(t)}=\frac{\dot{b}(t)}{b(t)} - \frac{\dot{p}(t)}{p(t)} = \frac{\dot{b}(t)}{b(t)} - \pi(t).
\end{equation*}
Thus, the real stock of bonds evolves according to
\begin{equation*}
\dot{w}(t)=\frac{\dot{b}(t)}{p(t)} - \pi(t) \cdot w(t).
\end{equation*}
Using the value of $\dot{b}(t)$ given by equation~\eqref{e:dotb}, we obtain the following flow budget constraint:
\begin{equation}
\dot{w}(t)= r(t) w(t) + \bs{1-u(t)} a l  - \bs{1+\tau(\t(t))}  c(t) - \frac{T(t)}{p(t)}.
\label{e:dotw}\end{equation}

\paragraph{Utility from services} The household consumes $c(t)$ services. From them, it enjoys flow utility
\begin{equation*}
\frac{\s}{\s-1}  c(t)^{(\s-1)/\s},
\end{equation*}
where $\s>1$ governs the concavity of the utility function.

\paragraph{Utility from relative wealth} Because government bonds are the only store of wealth, the household's real wealth is the real stock of bonds that it holds, $w(t)$. The household's relative real wealth is $w(t)-\bar{w}(t)$, where $\bar{w}(t)$ is average real wealth across all households. From its relative real wealth, the household enjoys flow utility
\begin{equation*}
x(w(t)-\bar{w}(t)), 
\end{equation*}
where the function $x: \R \to \R$ is increasing and strictly concave.

\paragraph{Household's problem} The problem of the household is to choose time paths for $c(t)$ and $w(t)$ to maximize the discounted sum of flow utilities
\begin{equation}
\int_{0}^{\infty}e^{-\d t} \bs{\frac{\s}{\s-1}  c(t)^{(\s-1)/\s}+x(w(t)-\bar{w}(t))}dt,
\label{e:utility}\end{equation}
where $\d>0$ is the time discount rate. The household is subject to the budget constraint given by equation~\eqref{e:dotw} and to a borrowing constraint preventing Ponzi schemes. The household takes as given the paths of $\t(t)$, $u(t)$, $p(t)$, $i(t)$, $T(t)$, and $\bar{w}(t)$; and initial real wealth $w(0)$.

\paragraph{Hamiltonian} To solve the household's problem, we set up the current-value Hamiltonian:
\begin{align*}
\Hc(t,c(t),w(t))&= \frac{\s}{\s-1}  c(t)^{(\s-1)/\s}+x(w(t)-\bar{w}(t)) \\
&+\g(t) \bs{r(t) w(t) + \bs{1-u(t)} al - \bs{1+\tau(\t(t))} c(t) - \frac{T(t)}{p(t)}},
\end{align*}
with control variable $c(t)$, state variable $w(t)$, and costate variable $\g(t)$.

\paragraph{Euler equation} The necessary conditions for an interior solution to the maximization problem are $\pdx{\Hc}{c}=0$, $\pdx{\Hc}{w}=\d \g(t) -\dot{\g}(t)$, and the appropriate transversality condition \cp[theorem~7.13]{A09}. In fact, since the utility function is strictly concave, interior paths of $c(t)$ and $w(t)$ that satisfy these conditions would constitute the unique global maximum of the household's problem \cp[theorem~7.14]{A09}.

The necessary conditions give
\begin{align}
c(t)^{-1/\s} &= \g(t) \bs{1+\tau(\t(t))}\label{e:euler1}\\
\dot{\g}(t)&= \bs{\d - r(t)} \g(t) - x'(w(t)-\bar{w}(t)).\label{e:euler2}
\end{align}
These two equations form the basis of the consumption Euler equation. Without recruiting cost ($\k=0$ so $\tau=0$) and no utility from wealth ($x'=0$), the equations reduce to the standard continuous-time Euler equation: 
\begin{equation*}
\frac{\dot{c}(t)}{c(t)} =\s [r(t)-\d].
\end{equation*}

\subsection{Price of services}

Because search is random on the service market, producers cannot attract customers by lowering their prices, and customers cannot attract producers by offering higher prices. Instead, each producer-customer pair agrees on a price in a situation of bilateral monopoly. The situation arises because when a producer and a customer match, a positive surplus is generated. Indeed, the producer prefers working for the customer than waiting to match with another customer because unemployment is not as productive as employment. And the customer prefers hiring the producer than hiring somebody else because recruiting is costly.

To resolve the bilateral monopoly, we specify a simple price mechanism whereby the price $p(t)$ grows at a constant rate of inflation $\pi$:
\begin{equation}
p(t) = e^{\pi t},
\label{e:p}\end{equation}
after normalizing $p(0)=1$. Following \ct{H05}, we interpret this price mechanism as a social norm. In the model the inflation norm responds neither to monetary policy nor to unemployment; but in reality the norm could be shaped over time by communication from the central bank.

\subsection{Monetary policy}

The central bank simply follows an interest-rate peg: it maintains the normal interest rate to 
\begin{equation}
i(t) = i.
\label{e:i}\end{equation}
Because of the ZLB, the nominal interest rate cannot be negative: $i \geq 0$. Since the inflation rate and nominal interest rate are fixed, the real interest rate is fixed to 
\begin{equation}
r = i - \pi.
\label{e:r}\end{equation}
We impose $r<\d$ so the model has a solution.

\subsection{Solution of the model}

We now solve the model.

\paragraph{Dynamics} The dynamics of the model are governed by equation~\eqref{e:euler2}, which is an Euler equation. Given the price mechanism and monetary policy, and given that all households are identical, this equation can be simplified. First, the real interest rate is fixed to $r$ at all times. Second, all households save the same, so their relative wealth is $0$ at all times. Hence equation~\eqref{e:euler2} reduces to
\begin{equation}
\dot{\g}(t) = \bs{\d - r} \g(t) - x'(0).
\label{e:euler}\end{equation}

\begin{figure}[p]
\includegraphics[scale=0.25,page=1]{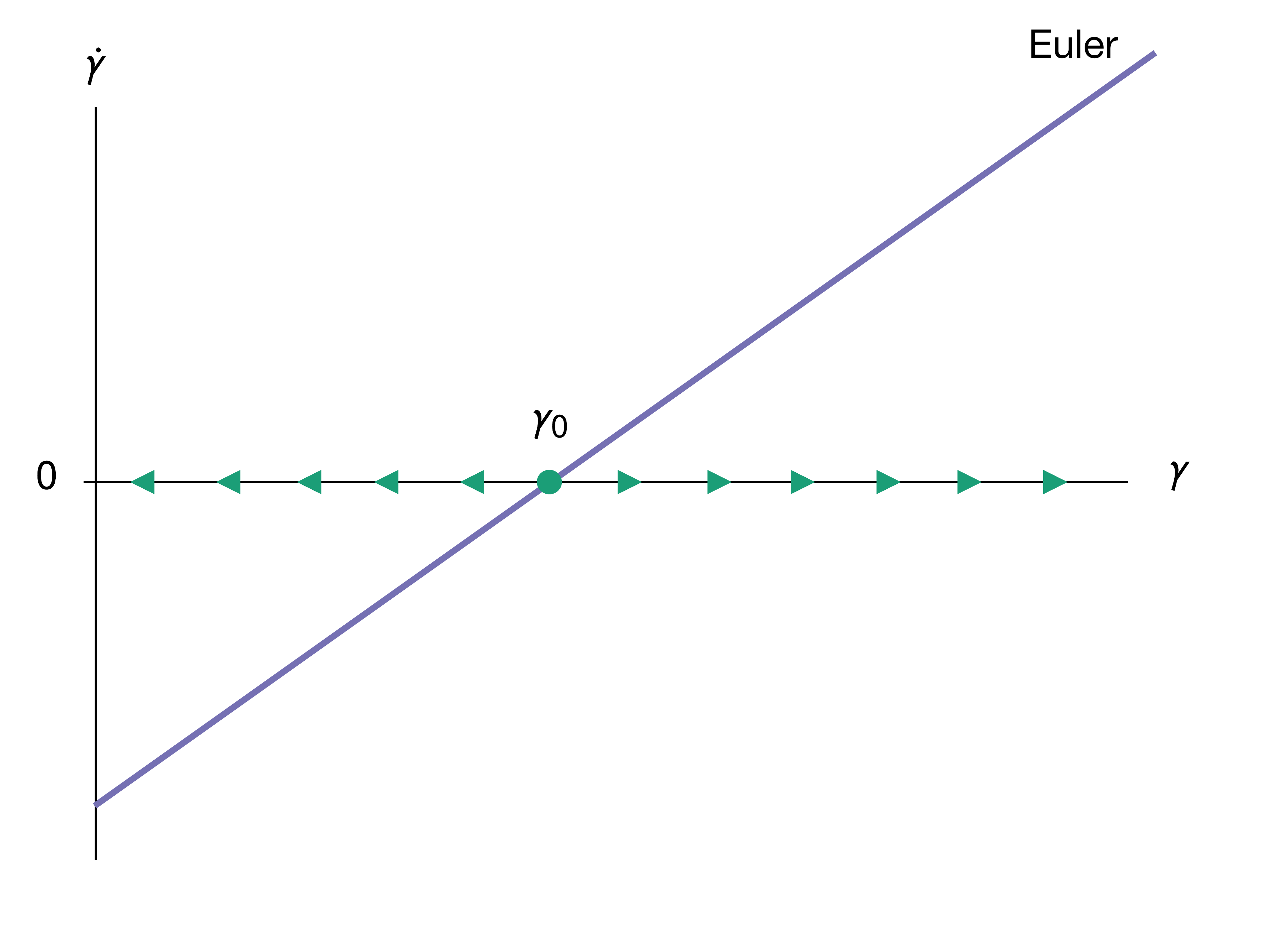}
\caption{Phase line of the Euler equation}
\note[Notes: ]{The Euler equation is given by equation~\eqref{e:euler}. The variable $\g$ is the costate variable associated with the household's real wealth. The critical point $\g_0$ is given by equation~\eqref{e:gamma}.}
\label{f:euler}\end{figure}

This Euler equation is an autonomous, first-order, and linear differential equation. It can be analysed using a phase line (figure~\ref{f:euler}). Since $r<\d$, it admits a unique critical point:
\begin{equation}
\g_0 =   \frac{x'(0)}{\d - r}.
\label{e:gamma}\end{equation}
Furthermore, the critical point is a source: $\dot{\g}>0$ when $\g>\g_0$, and  $\dot{\g}<0$ when $\g<\g_0$.

The costate variable $\g(t)$ is nonpredetermined at time $t$. One solution of the Euler equation is therefore constant over time: $\g$ jumps to $\g_0$ at time $0$ and remains at that point thereafter. (If $\g$ jumps to another position at time $0$, it diverges to $-\infty$ or $+\infty$ as $t\to \infty$.) Equation~\eqref{e:euler1} indicates that when $\g(t)=\g_0$, consumption is 
\begin{equation}
c(t) = \bs{\frac{\d - r}{x'(0)}\cdot \frac{1}{1+\tau(\t(t))}}^{\s}.
\label{e:cd}\end{equation}
Because an interior path of $c(t)$ that satisfies the optimality conditions is the unique global maximum of the household's problem, the consumption path given by equation~\eqref{e:cd} is the unique global maximum of the household's problem and thus the unique relevant solution of the Euler equation.

\paragraph{Structure of the solution} The Euler equation does not induce any dynamics because the costate variable $\g$ jumps to the critical value $\g_0$ at time $0$. Thus, the model can be solved from a set of static equations. The first step is to determine output $y$ and tightness $\t$. The other variables are then computed from $y$ and $\t$.

\paragraph{Aggregate supply} Combining equations~\eqref{e:y} and~\eqref{e:u}, we obtain 
\begin{equation}
y = \frac{f(\t)}{\l+f(\t)}  a l.
\label{e:as}\end{equation} 
We refer to this relationship between output and tightness as the aggregate supply (AS) curve. It gives the amount of output produced when unemployment is on the Beveridge curve. 

\begin{figure}[p]
\includegraphics[scale=0.25,page=2]{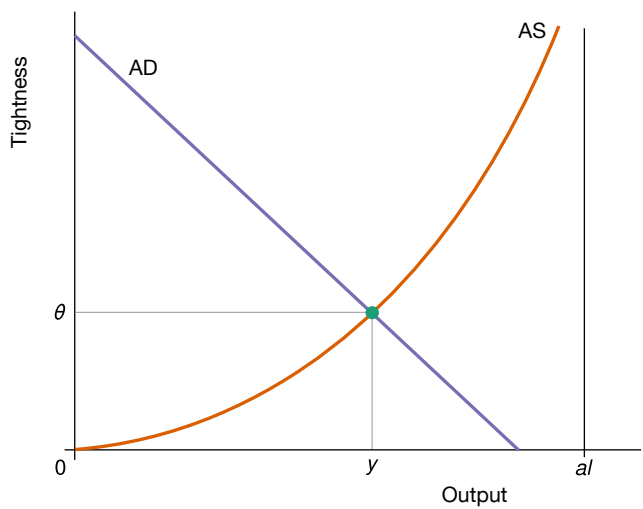}
\caption{Solution of the model}
\note[Notes: ]{The AS curve is given by equation~\eqref{e:as}. The AD curve is given by equation~\eqref{e:ad}. The variable $a$ is labour productivity; $l$ is labour-force size; and $al$ is aggregate productive capacity. The intersection of the AD and AS curves gives the output $y$ and tightness $\t$ that solve the model. The other variables are computed from $y$ and~$\t$.}
\label{f:solution}\end{figure}

The AS curve is displayed in figure~\ref{f:solution}. It starts at $y=0$ when $\t=0$, because $f(0)=0$. It then increases in $\t$, because $f(\t)$ increases in $\t$ and $\l>0$. It finally asymptotes to $y = al$ when $\t\to \infty$, because $f(\t)\to \infty$ when $\t\to \infty$. Output is increasing in tightness along the AS curve because when tightness is higher, it is easier to find work, so the Beveridgean unemployment rate is lower, which means that more workers are employed and producing services. 

\paragraph{Aggregate demand} Combining equations~\eqref{e:c} and~\eqref{e:cd}, we get
\begin{equation}
y = \bs{\frac{\d - r}{x'(0)}}^{\s} \cdot \frac{1}{[1+\tau(\t)]^{\s-1}}.
\label{e:ad}\end{equation}
We refer to this relationship between output and tightness as the aggregate demand (AD) curve. It is the amount of output demanded by households when consumption satisfies the Euler equation, and therefore when households optimally save over time. 

The AD curve is displayed in figure~\ref{f:solution}. It starts at $y=[(\d - r)/x'(0)]^{\s}$ when $\t=0$, because $\tau(0)=0$. It then decreases in $\t$, because $\tau(\t)$ increases in $\t$ and $\s>1$. It finally reaches $y = 0$ when $\t=\t_{\tau}$, because $\tau(\t)\to \infty$ when $\t\to \t_{\tau}$. Output is decreasing in tightness along the AD curve because when tightness is higher, it is harder to recruit employees, so consuming is less desirable relative to accumulating wealth.

\paragraph{Output and tightness} To solve the model, output $y$ and tightness $\t$ must satisfy both the AS curve and the AD curve. Hence, the model's solution is given by the intersection of the AS and AD curves (figure~\ref{f:solution}). Since the intersection always exists and is unique, the model admits a unique solution.

\paragraph{Other variables} The other variables are computed using output $y$ and tightness $\t$. The unemployment rate $u$ is computed from the Beveridge curve, given by equation~\eqref{e:u}. Employment $n$ is computed from the labour-force constraint, given by equation~\eqref{e:n}. The number of vacancies $v$ is computed from equation~\eqref{e:theta}. Consumption $c$ is computed from equation~\eqref{e:c}. The price level $p(t)$ is given by the price norm, described by equation~\eqref{e:p}. The inflation rate $\pi$ is fixed and given by the price norm. The nominal interest rate $i$ is fixed and given by monetary policy. The real interest rate $r$ is given by equation~\eqref{e:r}. The lump-sum tax $T(t)$ is determined by fiscal policy. Real wealth $w(t)$ follows equation~\eqref{e:dotw}. The nominal stock of bonds $b(t)$ is given by equation~\eqref{e:w}. 

\subsection{Sources of unemployment}

Adapting the work of \ct{M09}, we isolate the Keynesian and frictional components of unemployment. The Keynesian components arises because of deficient aggregate demand, and the frictional components because of the matching process. 

\paragraph{Keynesian unemployment} Without recruiting costs ($\k=0$ so $\tau=0$), output is determined by the AD curve and equals $y_k = \min\bc{al,[(\d - r)/x'(0)]^{\s}}$. The corresponding unemployment rate is $u_k = 1-y_k/al$, which we call Keynesian unemployment. Keynesian unemployment is positive whenever aggregate demand is weak ($x'(0) > (\d-r) (al)^{-1/\s}$). In this case, it satisfies
\begin{equation*}
u_{k} = 1- \frac{1}{al} \bs{\frac{\d - r}{x'(0)}}^{\s}.
\end{equation*}
Keynesian unemployment is larger when aggregate demand is more depressed. This happens when the discount rate is lower, the real interest rate is larger, or the marginal utility of wealth is larger.

\paragraph{Frictional unemployment} Because recruiting costs are positive, actual unemployment is higher than Keynesian unemployment. Frictional unemployment measures additional unemployment attributable to recruiting costs: 
\begin{equation*}
u_f = u - u_k.
\end{equation*}

\section{Inefficiency}

We define and characterize the efficient unemployment rate. We find that in the model the rate of unemployment is generally inefficient.

\subsection{Social welfare}

The solution of the model is constant over time, so social welfare is adequately measured by its flow value. To construct flow social welfare, we use the flow utility appearing in equation~\eqref{e:utility}; the labour-force constraint, given by equation~\eqref{e:n}; the production function, given by equation~\eqref{e:y}; the fact that $\k  v$ employed workers are devoted to recruiting; and the fact that relative wealth is $0$. We find that flow welfare is a function of the unemployment rate $u$ and vacancy level $v$ given by 
\begin{equation}
\frac{\s}{\s-1} \bc{a [(1-u)l - \k v]}^{(\s-1)/\s}+x(0).
\label{e:sw}\end{equation}

\subsection{Definition of efficiency}

The efficient unemployment rate and vacancy level, denoted $u^*$ and $v^*$, are the solution to the problem of a planner who chooses unemployment and vacancies to maximize social welfare, given by equation~\eqref{e:sw}, subject to the Beveridge curve, given by equation~\eqref{e:u}. The Beveridge curve can be expressed as
\begin{equation}
v = \bs{\frac{\l(1-u)}{\m u^{\h}}}^{1/(1-\h)}  l.
\label{e:v}\end{equation}
Hence, the efficient unemployment rate maximizes 
\begin{equation}
(1-u)l - \k v(u),
\label{e:wu}\end{equation}
where $v(u)$ is defined by equation~\eqref{e:v}. The efficient tightness is $\t^* = v^*/(u^* l)$. 

\subsection{Efficient tightness and unemployment rate}

We characterize the efficient tightness before turning to the efficient unemployment rate.

\paragraph{Sufficient-statistic formula} According to the sufficient-statistic formula derived by \ct[proposition~2]{MS16}, here the efficient tightness satisfies
\begin{equation}
\t^* = \frac{1}{\k \e}.
\label{e:ms16}\end{equation}
The statistic $\e$ is the Beveridge elasticity: the elasticity of vacancies with respect to unemployment in equation~\eqref{e:v}, normalized to be positive.

It is actually easy to rederive equation~\eqref{e:ms16} in this model. Since $u^*$ maximizes welfare, and welfare is given by equation~\eqref{e:wu}, the following first-order condition holds: 
\begin{equation*}
- \k v'(u^*) = l. 
\end{equation*}
This condition is equivalent to 
\begin{equation*}
\k \bs{ \frac{- u^* v'(u^*)}{v^*}} = \frac{u^* l}{v^*},
\end{equation*}
 which gives $\k\e=1/\t^*$ and then equation~\eqref{e:ms16}.

\paragraph{Structural formula} To learn more about the efficient tightness, we compute the Beveridge elasticity and transform equation~\eqref{e:ms16} into a structural formula. The Beveridge elasticity satisfies
\begin{equation*}
\e = -\odl{v}{u} = \frac{1}{1-\h}\bp{\h+\frac{u}{1-u}}.
\end{equation*}
Using equation~\eqref{e:u}, we express the elasticity as a function of tightness:
\begin{equation}
\e = \frac{1}{1-\h}\bs{\h+\frac{\l}{f(\t)}}.
\label{e:e}\end{equation}
Combining equations~\eqref{e:ms16} and \eqref{e:e}, we find that the efficient tightness is implicitly defined by the following structural formula:
\begin{equation}
\frac{\k}{1-\h}  \bs{\h\t^* + \frac{\l}{q(\t^*)}} = 1.
\label{e:thetastar}\end{equation}
The efficient tightness is well defined because the equation admits a unique solution on $(0,\t_{\tau})$.\footnote{The left-hand side of equation~\eqref{e:thetastar} is a continuous function of $\t^*$. When $\t^*$ increases from $0$ to $\t_{\tau}$, it strictly increases from $0$ to 
\begin{equation*}
\frac{\k\h\t_{\tau}}{1-\h} + \frac{\k\l}{1-\h} \cdot \frac{1}{q(\t_{\tau})} =  \frac{1+\k\h\t_{\tau}}{1-\h}>1.
\end{equation*}
Accordingly, equation~\eqref{e:thetastar} admits a unique solution on $(0,\t_{\tau})$.}

Equation~\eqref{e:thetastar} only involves parameters related to the matching process: the matching elasticity ($\h$), the recruiting cost ($\k$), the job-separation rate ($\l$), and the matching efficacy ($\m$). Thus, the matching process alone determines the efficient tightness.

\paragraph{Efficient unemployment rate} The efficient unemployment rate $u^*$ is computed from the efficient tightness $\t^*$, which is given by equation~\eqref{e:thetastar}, and the Beveridge curve, given by equation~\eqref{e:u}. Since the parameters in the Beveridge curve ($\l$, $\h$, $\m$) also enter equation~\eqref{e:thetastar}, the efficient unemployment rate is determined by the same parameters as the efficient tightness.

\subsection{Aggregate inefficiency} 

Tightness is determined by the AD and AS curves (figure~\ref{f:solution}). In turn, the AD curve is determined by the real interest rate ($r$), discount rate ($\d$), and marginal utility of wealth ($x'(0)$); and the AS curve is determined by the labour productivity ($a$) and labour-force size ($l$). Because these parameters do not influence efficient tightness, actual and efficient tightnesses do not generally overlap. Therefore, the model economy is generally inefficient.

To measure how far the economy is from efficiency, we introduce the unemployment gap $u-u^{*}$. When the unemployment gap is positive, the unemployment rate is inefficiently high and tightness is inefficiently low. Conversely, when the unemployment gap is negative, the unemployment rate is inefficiently low and tightness is inefficiently high.

\subsection{Bilateral efficiency}

Finally, we show that despite the model's aggregate inefficiency, relationships between producers and consumers are bilaterally efficient.

\paragraph{Consumer surplus} When a consumer matches with a producer, she strictly prefers purchasing a service at price $p$ than not purchasing it. This is because the consumer anticipated to purchase the service at that price. She matched with a producer only because it was beneficial to her. Formally, households choose their consumption such that equation~\eqref{e:euler1} holds: the utility from consuming 1 service ($c^{-1/\s}$) equals the value of 1 unit of real wealth ($\g$) times the real price of the service inclusive of the recruiting cost ($1 + \tau(\t)$). Therefore, the utility from consuming 1 service is strictly larger than the value that can be obtained by not trading and saving $p$ units of money, or $1$ unit of real wealth ($c^{-1/\s} > \g$). 

The consumer could also refuse to hire the current producer and hire somebody else instead. She would be worse-off, however, because the other producers charge the same price, but recruiting a new producer is costly.

\paragraph{Producer surplus} When a producer is matched with a consumer, she strictly prefers selling services at price $p$ than remaining unemployed. This is because an unemployed worker is busy searching for a job, so she cannot contribute to the household's well-being.  

The producer could also refuse to work for the current consumer and try to work for somebody else instead. She would be worse-off, however, because the other consumers pay the same price, but finding a new consumer takes time---during which the jobseeker is not earning anything.

\paragraph{Summary} Both producers and consumers enjoy some surplus. They prefer transacting at the price given by equation~\eqref{e:p} than not transacting or transacting with somebody else. Hence, although inflation is fixed and not determined by some market or bargaining mechanism, prices are bilaterally efficient. By contrast, in non-matching models, fixed prices create bilateral inefficiencies \cp{B77}.

\section{Business cycles}\label{s:cycles}

\begin{figure}[p]
\includegraphics[scale=0.25,page=3]{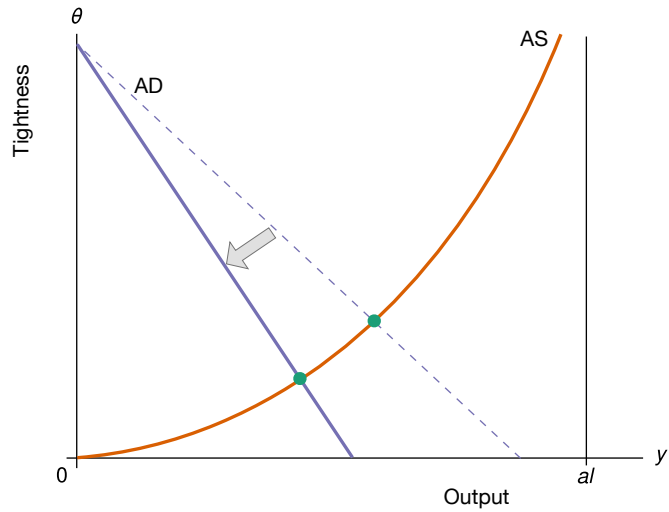}
\caption{Response to a negative aggregate demand shock}
\note[Notes: ]{The AS and AD curves are constructed in figure~\ref{f:solution}. The shock is an unexpected permanent decrease in the discount rate ($\d$) or an unexpected permanent increase in the marginal utility of wealth ($x'(0)$). The graph shows the response of output and tightness to the shock; table~\ref{t:cycles} describes the response of other variables.}
\label{f:ad}\end{figure}

\begin{figure}[p]
\includegraphics[scale=0.25,page=4]{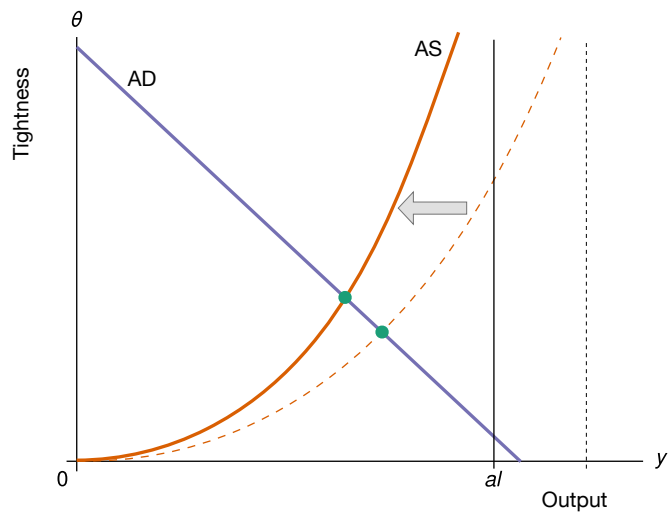}
\caption{Response to a negative aggregate supply shock}
\note[Notes: ]{The AS and AD curves are constructed in figure~\ref{f:solution}. The shock is an unexpected permanent decrease in labour productivity ($a$) or in labour-force size ($l$). The graph shows the response of output and tightness to the shock; table~\ref{t:cycles} describes the response of other variables.}
\label{f:as}\end{figure}

We describe how the model economy responds to business-cycle shocks---both aggregate demand and aggregate supply shocks. As we consider unexpected permanent shocks, we derive the response by comparative statics. The results are summarized in table~\ref{t:cycles} and illustrated in figures~\ref{f:ad} and \ref{f:as}.

\begin{table}[t]
\caption{Effects of business-cycle and policy shocks}
\begin{tabular*}{\textwidth}{@{\extracolsep{\fill}}l*{5}{c}}
\toprule
& 		 & 			& 		 &\multicolumn{2}{c}{Unemployment rate}\\
\cmidrule{5-6}
& Tightness & Output & Employment & Actual  & Efficient\\
\midrule
\textbf{Aggregate demand shocks} 	&  &  &  & &  \\
Decrease in discount rate	& $-$ & $-$  & $-$ & $+$ & $0$ \\
Increase in marginal utility of wealth & $-$ & $-$  & $-$ & $+$ & $0$ \\
\midrule
\textbf{Aggregate supply shocks}   & &  &  & &  \\
Decrease in labour productivity & $+$  & $-$  & $+$ & $-$ & $0$ \\
Decrease in labour-force size & $+$  & $-$  & $-$ & $-$ & $0$ \\
\midrule
\textbf{Policy shocks} & &  &  &  \\
Decrease in nominal interest rate & $+$ & $+$  &  $+$  &  $-$ & $0$ \\
Increase in wealth tax rate	& $+$ & $+$  &  $+$  &  $-$ & $0$ \\
\bottomrule
\end{tabular*}
\note[Notes: ]{The table reports the effects of unexpected permanent changes in aggregate demand, aggregate supply, and policy on key variables of the model. The response of tightness and output is obtained by comparative statics in figures~\ref{f:ad}--\ref{f:tax}. The response of employment and the actual unemployment rate then follows from equations \eqref{e:n}, \eqref{e:y}, and \eqref{e:u}. The response of the efficient unemployment rate derives from equations \eqref{e:u} and \eqref{e:thetastar}.}
\label{t:cycles}\end{table}

\subsection{Aggregate demand shocks}

We first examine aggregate demand shocks: either a shock to the discount rate ($\d$) or to the marginal utility of wealth ($x'(0)$). For concreteness, we consider a decrease in the discount rate or an increase in the marginal utility of wealth. After such shock, households become more thrifty: they desire to save more and consume less, which depresses aggregate demand (equation~\eqref{e:ad}). The AD curve rotates inward, so  output and tightness fall (figure~\ref{f:ad}). Since tightness falls, the unemployment rate rises (equation~\eqref{e:u}), and the employment level drops (equation~\eqref{e:n}). Since efficient tightness and efficient unemployment rate are unaffected by aggregate demand shocks (equations~\eqref{e:thetastar} and \eqref{e:u}), the unemployment gap increases.

The Keynesian paradox of thrift emerges here. When the marginal utility of wealth is higher, people want to increase their wealth relative to their peers, so they favour saving over consumption. But in the aggregate relative wealth is fixed at 0 because everybody is the same; the only way to increase saving relative to consumption is to consume less.

Finally, the results are the same whether the economy is at the ZLB ($i=0$) or not ($i>0$). 

\subsection{Aggregate supply shocks}

Next, we examine aggregate supply shocks: either a shock to labour productivity ($a$) or to labour-force size ($l$). For concreteness, we consider a decrease in labour productivity or in labour-force size. If workers are less productive, or if fewer workers participate in the labour force, then mechanically the aggregate supply is depressed (equation~\eqref{e:as}).  The AS curve shifts inward, so output falls while tightness rises (figure~\ref{f:as}). Since tightness rises, the unemployment rate decreases (equation~\eqref{e:u}). Since efficient tightness and efficient unemployment rate are unaffected by aggregate supply shocks (equations~\eqref{e:thetastar} and \eqref{e:u}), the unemployment gap decreases.

The response of the employment level depends on the shock. When the labour-force size shrinks, employment drops because output drops and $n = y/a$ (equation~\eqref{e:y}). On the other hand, when labour productivity falls, employment rises because unemployment falls and $n = (1-u) l$ (equation~\eqref{e:n}). The prediction that a decrease in labour productivity leads to lower output but higher employment conforms to US evidence \cp{GR04,BFK06,FR09}.

Once again, the results are the same whether the economy is at the ZLB or not. The prediction that aggregate supply shocks have the same effects at the ZLB and away from it accords with the evidence presented by \ct{W13}.

\subsection{Sources of business cycles}

The comovements of output and inflation are commonly used to identify the source of business-cycle fluctuations \cp{H00,GZ19}. In this model, inflation is fixed, so the approach is not available. Instead, the source of fluctuations can be identified from the comovements of output and unemployment rate. Indeed, all aggregate demand shocks generate negative comovements between output and unemployment rate, whereas aggregate supply shocks generate positive comovements (table~\ref{t:cycles}).

In the United States and other developed economies, output and unemployment rate are negatively correlated in the short run---as stipulated by Okun's law \cp{BLL13}. According to the model, such negative comovements indicate that business-cycle fluctuations are mostly caused by aggregate demand shocks.

\section{Monetary policy}\label{s:monetary}

We now show how monetary policy can stimulate aggregate demand. We also show that it is optimal for monetary policy to maintain unemployment at its efficient level. Hence, the optimal nominal interest rate can be computed from two sufficient statistics: the prevailing unemployment gap and the monetary multiplier. Last, we show that at the ZLB  monetary policy can be substituted by a wealth tax.

\begin{figure}[p]
\includegraphics[scale=0.25,page=5]{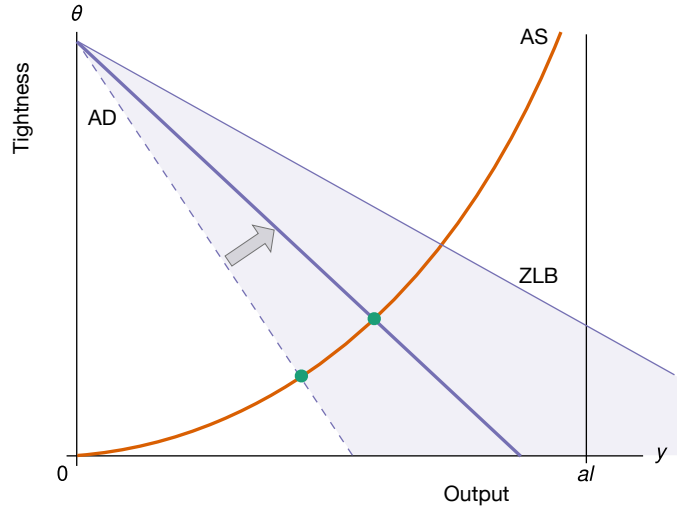}
\caption{Response to a decrease in nominal interest rate}
\note[Notes: ]{The AS and AD curves are constructed in figure~\ref{f:solution}. The purple cone indicates all the positions that the AD curve can reach when the nominal interest rate is reduced from its current level. The ZLB curve shows the position of the AD curve at the ZLB; it is the most outward position that the AD curve can reach; it is  obtained from equation~\eqref{e:ad} when $r = -\pi$. The graph shows the response of output and tightness to an unexpected permanent decrease in the nominal interest rate ($i$); table~\ref{t:cycles} describes the response of other variables.}
\label{f:monetary}\end{figure}

\begin{figure}[p]
\includegraphics[scale=0.25,page=6]{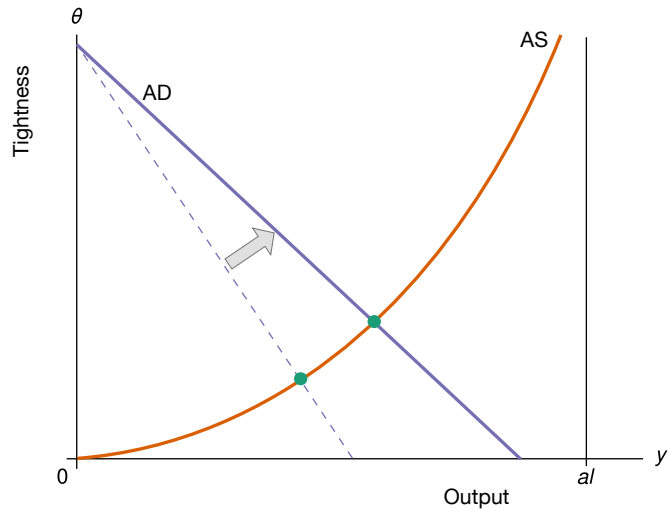}
\caption{Response to an increase in wealth tax rate}
\note[Notes: ]{The AS curve is constructed in figure~\ref{f:solution}. The AD curve is given by equation~\eqref{e:adt}. The graph shows the response of output and tightness to an unexpected permanent increase in the wealth tax rate ($\tau_{w}$); table~\ref{t:cycles} describes the response of other variables.}
\label{f:tax}\end{figure}

\subsection{Monetary policy shock}

We start by describing how the model economy responds to a monetary policy shock. We consider an unexpected permanent decrease in the nominal interest rate ($i$). As inflation ($\pi$) remains constant, the real interest rate ($r = i-\pi$) falls. When the real rate falls, wealth has lower returns, so households desire to save less and consume more, which boosts aggregate demand (equation~\eqref{e:ad}). The AD curve rotates outward, so output and tightness increase (figure~\ref{f:monetary}). Since tightness rises, the unemployment rate decreases (equation~\eqref{e:u}), and the employment level increases (equation~\eqref{e:n}). Since efficient tightness and efficient unemployment rate are unaffected by monetary policy (equations~\eqref{e:thetastar} and \eqref{e:u}), the unemployment gap decreases.

\subsection{Optimal monetary policy}

\begin{figure}[p]
\includegraphics[scale=0.25,page=7]{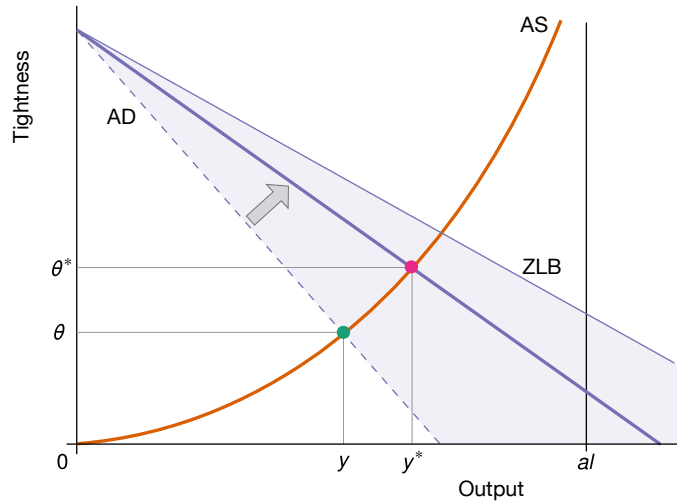}
\caption{Small unemployment gap: optimal monetary policy restores efficient unemployment}
\note[Notes: ]{The AS and AD curves are constructed in figure~\ref{f:solution}. The ZLB curve is constructed in figure~\ref{f:monetary}. The tightness $\t^*$ and output $y^*$ constitute the efficient allocation. Initially, aggregate demand is given by the dashed AD curve, and tightness is $\t$. Since $\t<\t^*$, the initial unemployment gap is positive. It is optimal for monetary policy to eliminate the unemployment gap, so the interest rate should drop to stimulate aggregate demand and bring tightness to $\t^*$. Since the efficient allocation is in the purple cone, optimal monetary policy can be implemented.}
\label{f:small}\end{figure}

\paragraph{Description of the optimal policy} Social welfare is maximized when the rate of unemployment is efficient---that is, when the unemployment gap is 0. Since monetary policy modulates the AD curve without creating any distortions, the optimal monetary policy is to target the efficient unemployment rate $u^*$ so as to eliminate the unemployment gap. Formally, the optimal nominal interest rate $i^*$ must be set so that
\begin{equation}
 u(i^*)=u^*,
\label{e:ustar}\end{equation} 
where $u(i)$ is the unemployment rate given by the model when the nominal interest rate is $i$. Due to the ZLB, such policy is only feasible if $i^* \geq 0$. The optimal monetary policy is illustrated in figure~\ref{f:small}.

\paragraph{Sufficient-statistic formula} Starting from a nominal interest rate $i>0$ and an inefficient unemployment rate $u \neq u^*$, what should the central bank do to bring the unemployment rate to $u^*$? A first-order expansion of $u(i)$ around $i^*$ gives 
\begin{equation*}
u(i) = u^* + \od{u}{i} (i-i^*) + O((i-i^*)^2),
\end{equation*}
so up to a second-order term the optimal interest rate $i^*$ satisfies
\begin{equation}
 i - i^* =  \frac{u-u^*}{du/di}.
\label{e:monetary}\end{equation}
The statistic $i-i^*$ indicates the decrease in interest rate required to reach the optimal policy. The statistic $u-u^*$ is the prevailing unemployment gap. And the statistic $du/di>0$ is the monetary multiplier: the percentage-point decrease in unemployment achieved by lowering the nominal interest rate by 1 percentage point.

Optimal monetary policy depends on two sufficient statistics: unemployment gap and monetary multiplier. When the current unemployment gap is positive, the nominal interest rate must be reduced to reach efficiency. The nominal interest rate must be reduced more drastically when the unemployment gap is larger. The same is true when the monetary multiplier is smaller---that is, when monetary policy is less effective.

\paragraph{Application to other models} Equation~\eqref{e:monetary} is quite general. It applies to any model that is neo-Wicksellian, with a Beveridge curve, and with divine coincidence. In a neo-Wicksellian model, the central bank stabilizes the economy by adjusting the nominal interest rate \cp{W03}. In a model with a Beveridge curve, the efficient unemployment rate can be computed from estimable statistics \cp{MS16}. Last, under divine coincidence, inflation is guaranteed to be on target when the unemployment rate is efficient \cp{BG07}. In such model, the optimal monetary policy is given by equation~\eqref{e:ustar}, and therefore by equation~\eqref{e:monetary}. By bringing the unemployment rate to $u^*$, the central bank also brings inflation to its target.

\subsection{Implementing optimal monetary policy}

Because the formula is simple and based on estimable statistics, it should be easy to implement.

\paragraph{Unemployment gap} \ct{MS16} estimate the unemployment gap in the United States. Because the estimate is based on the Beveridge curve, it applies to this model, and it can be used to implement equation~\eqref{e:monetary}. They find that between 1951 and 2019, the unemployment gap is almost always positive and highly countercyclical, which indicates that the US economy is generally inefficiently slack, and especially so in slumps. Accordingly, the Federal Reserve has scope to stabilize the economy over the business cycle.

\paragraph{Monetary multiplier} A large literature estimates the effect of the federal funds rate on unemployment \cp{CEE99,R15}. \ct{C12} reviews existing estimates of the monetary multiplier and provides his own. Using a traditional VAR approach, \ct[p.~5]{C12} finds $du/di = 0.16$. This value accords with previous VAR and FAVAR estimates of the monetary multiplier, with the exception of the large estimate $du/di = 0.6$ obtained by \ct{BB92}. Using the narrative approach of \ct{RR04}, \ct[p.~7]{C12} obtains a much larger estimate: $du/di = 0.93$. To reconcile the difference between the two approaches, \ct[p.~8]{C12} proposes a hybrid approach that yields a medium estimate: $du/di = 0.40$. The hybrid approach appears robust: across numerous specifications, it yields monetary multipliers between $0.4$ and $0.6$ \cp[p.~12]{C12}. In sum, a midrange estimate of the US monetary multiplier seems to be $du/di = 0.5$.

\paragraph{Optimal response to unemployment fluctuations} Combined with a monetary multiplier of $0.5$, equation~\eqref{e:monetary} says that the Federal Reserve should cut its interest rate by $1/0.5 = 2$ percentage points for each percentage point of unemployment gap. For instance, at the peak of the Great Recession, the US unemployment gap was above $5$ percentage points \cp[figure~6]{MS16}. So the Federal Reserve needed to reduce the federal funds rate by more than $2 \times 5 =10$ percentage points to eliminate the unemployment gap. Since the funds rate was about $5\%$ at the beginning of the recession, the Fed needed to set a negative rate to eliminate the unemployment gap---which it could not do because of the ZLB. This explains why the ZLB rapidly became binding in the Great Recession, and why unemployment remained high throughout it.

\paragraph{Evaluating the Federal Reserve's behaviour} Our calibrated formula implies that the Fed should reduce the federal funds rate by $2$ percentage points for any one-percentage-point increase in unemployment gap. In our model, the efficient unemployment rate is constant over the business cycle, so the unemployment gap moves one-for-one with the unemployment rate. Hence, the Fed should reduce its interest rate by $2$ percentage points for each one-percentage-point increase in unemployment. 

This is roughly what the Fed appears to do. \ct[figure~2]{BB92} find that an unemployment increase by $18$ basis points leads to a reduction in the funds rate by $28$ basis points. Thus, the Fed raises the funds rate by $28/18 = 1.6$ percentage points for each one-percentage-point increase in unemployment. Similarly, \ct[figure~1]{SW01} find that after a surprise increase in unemployment, the funds rate decreases. On impact, unemployment increases by $1$ percentage point and the funds rate drops by $2$ percentage points. After a year, unemployment has increased by $1.5$ percentage points and the funds rate has dropped by $3.2$ percentage points; that is, the funds rate drops by $3.2/1.5 = 2.1$ percentage points for each one-percentage-point increase in unemployment.
 
\begin{figure}[p]
\includegraphics[scale=0.25,page=8]{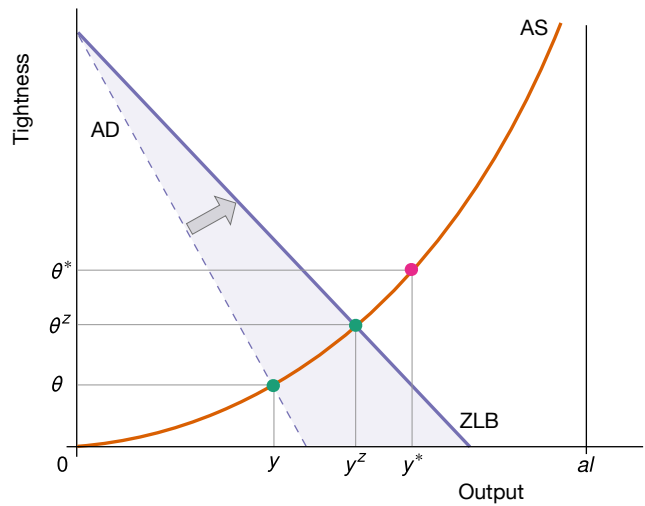}
\caption{Large unemployment gap: optimal monetary policy reaches the ZLB}
\note[Notes: ]{The AS, AD, and ZLB curves are constructed as in figure~\ref{f:small}, but here initial aggregate demand is more depressed, so the AD and ZLB curves are further inward. Accordingly, the initial unemployment gap is larger, and the efficient allocation falls outside of the purple cone. Hence, monetary policy cannot eliminate the unemployment gap before reaching the ZLB. The best that monetary policy can do is reduce the nominal interest rate to $0$, and bring the economy to the ZLB. There tightness is $\t^z<\t^*$, so the unemployment gap remains positive.}
\label{f:large}\end{figure}

\subsection{ZLB episodes}

 If equation~\eqref{e:monetary} gives $i^*<0$, optimal monetary policy is constrained by the ZLB. The best that the central bank can do is set the nominal interest rate to $0$. This situation arises if aggregate demand is particularly depressed.

 By bringing the economy to the ZLB, the central bank stimulates aggregate demand as much as possible without making the nominal interest rate negative. However, such expansionary policy is not sufficient to eliminate the unemployment gap. As a result, unemployment remains inefficiently high (figure~\ref{f:large}). 

\subsection{Wealth tax}

Once the ZLB is reached, monetary policy cannot stimulate the economy. But as we discuss here, a wealth tax can. 

\paragraph{Wealth tax rate} The government sets a tax rate $\tau_{w} \geq 0$ on households' wealth. Since bonds are the only store of wealth, it is bond holdings that are taxed.\footnote{In a model with money, such as in the first version of the paper \cp{MS14}, money would also be part of the tax base. Indeed, lowering the nominal interest rate below 0 requires a negative interest rate on money balances---effectively a wealth tax \cp{AK15}.} The household's tax liability at time $t$ is $\tau_{w} \cdot b(t)$. The tax revenue is rebated to households through a lump-sum transfer.

\paragraph{Budget constraint} The wealth tax modifies the household's flow budget constraint, but nothing else. The budget constraint becomes
\begin{equation}
\dot{w}(t)= (r- \tau_{w}) w(t)  +  \bs{1-u(t)} a l - \bs{1+\tau(\t(t))} c(t) - \frac{T(t)}{p(t)}.
\label{e:dotwt}\end{equation}

\paragraph{Aggregate demand} The wealth tax reduces the real rate of return on wealth from $r$ to $r-\tau_{w}$ (equation~\eqref{e:dotwt}); it does not affect the model in any other way. Hence, we solve the model with wealth tax like the model without it, except that the AD curve becomes
\begin{equation}
y = \bs{\frac{\d - r + \tau_{w}}{x'(0)}}^{\s} \cdot \frac{1}{[1+\tau(\t)]^{\s-1}}.
\label{e:adt}\end{equation}

\paragraph{Equivalence between wealth tax and monetary policy} The wealth tax perfectly substitutes for monetary policy because an increase in the wealth tax rate has the same effect as a commensurate decrease in the real interest rate (equation~\eqref{e:adt}). This is especially useful at the ZLB because the wealth tax is not constrained by the ZLB while monetary policy is.

\paragraph{Optimal wealth tax} Following the same steps as with monetary policy, we derive a sufficient-statistic formula for the optimal wealth tax. We consider an initial  unemployment rate $u \neq u^*$ and wealth tax rate $\tau_{w}$, and we assume that monetary policy remains constant. Then the optimal wealth tax rate $\tau_{w}^*$ satisfies
\begin{equation}
\tau_{w}^* - \tau_{w} =  \frac{u-u^*}{-du/d\tau_{w}},
\label{e:wealth}\end{equation}
The statistic $-du/d\tau_{w}$ is the tax multiplier: the percentage-point decrease in unemployment achieved by raising the wealth tax rate by 1 percentage point. 

We have not found any estimates of the tax multiplier, but in the model the tax multiplier equals the monetary multiplier---because wealth tax and monetary policy are isomorphic. As a first step, we opt to calibrate the tax multiplier from the estimate of the monetary multiplier: $-du/d\tau_{w} = 0.5$. We infer that at the ZLB, the government should raise the wealth tax by $2$ percentage points for every percentage point of unemployment gap.

\section{Conclusion}\label{s:conclusion}

This paper proposes a new model that aims to improve upon the New Keynesian model. To conclude, we highlight the main differences between this model and the New Keynesian model, and between their policy prescriptions. We also discuss possible extensions of the model.

\subsection{Comparison with the New Keynesian model}

We briefly explain how our model differs from the New Keynesian model.

\paragraph{Aggregate demand} Because wealth enters the utility function, the household's Euler equation is modified. As a result, the AD curve is nondegenerate: output is a decreasing function of the real interest rate, as in the Keynesian IS curve. Thanks to this AD curve, the model accommodates long-lasting or permanent ZLB episodes. By contrast, in the New Keynesian model, the AD curve is degenerate: it imposes that the real interest rate equals the discount rate, without involving output. The degeneracy creates anomalies when ZLB episodes are long-lasting \cp{MS18}.

\paragraph{Pricing} In our model, inflation is fixed. Thanks to the matching framework, however, all trades are bilateral inefficient. In New Keynesian models, inflation is determined by monopolistic pricing under price-setting frictions; bilateral inefficiencies are common \cp{HR20}.

\paragraph{Aggregate supply} We replace the New Keynesian Phillips curve by an AS curve derived from the Beveridge curve. The Beveridge curve itself emerges from the matching process. Because of it, the model features unemployment. In the New Keynesian model, by contrast, unemployment is absent.

\paragraph{Business cycles} In our model, aggregate demand and supply shocks create fluctuations in unemployment but not in inflation. In the New Keynesian model, these shocks create fluctuations in inflation but not in unemployment.

\paragraph{ZLB episodes}  Our model behaves the same at the ZLB and away from it. In both situations, the AD and AS curves are identical, and business-cycle and policy shocks have identical effects. In the New Keynesian model, by contrast, the ZLB is `a topsy-turvy world, in which many of the usual rules of macroeconomics are stood on their head' \cp[p.~1472]{EK11}.

\subsection{Comparison with the New Keynesian policy prescriptions}

Because this model differs fundamentally from the New Keynesian model, their policy prescriptions are significantly different.

\paragraph{Determinacy} In the New Keynesian model, the central bank must adhere to an active monetary policy to ensure determinacy---to ensure that the model's solution is locally unique. In our model, indeterminacy is never a risk, so the central bank does not need to respond mechanically to inflation or unemployment. The central bank can follow an interest-rate peg without creating indeterminacy. An implication is that indeterminacy is not a risk at the ZLB.

In fact, our model's solution is globally unique under an interest-rate peg. In the New Keynesian model, in sharp contrast, there are two steady states under active monetary policy: a normal steady state that is determinate and a ZLB steady state that is indeterminate \cp{BSU01}.\footnote{Similarly, non-New-Keynesian models developed to produce a determinate ZLB steady state often have several steady states: a ZLB steady state and at least one full-employment steady state \cp{M15,EMR17}.}

\paragraph{Monetary-policy target} A key policy lesson derived from New Keynesian models is that the central bank should only target inflation. As noted by \ct[pp.~459--460]{SU08}, `Policy mistakes are committed when policymakers are unable to resist the temptation to respond to output fluctuations\dots It follows that sound monetary policy calls for sticking to the basics of responding to inflation alone.'  In our model the exact opposite is true: the central bank should only target unemployment. An appropriate unemployment target is the efficient unemployment rate estimated by \ct{MS16}.

\ct[p.~1523]{Mc99} argues in favour of robustness: `searching for a rule that works reasonably well in a variety of models.' Inflation targeting would be completely ineffective in our model, so such policy is not as robust as previously thought.

\paragraph{Optimal monetary-policy rule} In the New Keynesian model, the optimal  monetary-policy rule is not implementable because the natural rate of interest is unobservable. Instead, researchers optimize the slopes of fixed monetary-policy rules to approximate the optimal policy as closely as possible \cp{TW10,G15}. As the optimization is done through simulations, the results are specific to a model and a calibration.

In our model, the unemployment gap is measurable, so we obtain an optimal interest-rate formula that the central bank can implement. The formula applies to any model with a Beveridge curve---even models with inflation fluctuations, as long as the divine coincidence holds. With the formula, the central bank can optimally adjust its interest rate after a shock has affected unemployment.

\paragraph{Monetary multiplier} Many studies estimate the effect of the federal funds rate on unemployment. Yet, these estimates of the monetary multiplier are useless in the New Keynesian model. They cannot be used to design policy or calibrate the model. In our model, on the other hand, the monetary multiplier is critical to design policy because it enters the optimal monetary-policy formula. The formula is implementable only because numerous estimates of the monetary multiplier are available.

\paragraph{Policies at the ZLB} In the New Keynesian model, forward guidance has   incredibly strong effects at the ZLB, so it is a policy of choice for the central bank. In our model, forward guidance has much more subdued effects, and it becomes impotent when the ZLB lasts long enough---for the reasons presented in \ct{MS18}. Instead, the government can use a wealth tax at the ZLB. By increasing the wealth-tax rate, the government stimulates aggregate demand and can eliminate the unemployment gap.

\subsection{Possible extensions}

Since our model is extremely stylized, it could be fruitfully extended along several dimensions. The extended model would be less economical, but it would be more amenable to quantitative analysis, and it would allow researchers to address other policy questions.

\paragraph{Endogenous inflation} Our model assumes that inflation is fixed. The justification is that producers strictly adhere to the norm of increasing their prices at a constant rate. The assumption seems valid to describe modern business cycles, with their low and stable inflation. But it is inappropriate to describe inflation dynamics in the medium run, when the inflation norm could be eroded by shocks or policy. It would therefore be valuable to develop a theory of endogenous inflation. An advantage of the matching framework is that it allows for a broad range of price-setting theories. It may therefore be possible to develop a theory of inflation consistent with the micro evidence that firms keep their prices stable by fear of antagonizing customers, who find certain price changes unfair \cp{BCL98}. \ct{EMM20} develop such a theory in a monopoly model; it might be possible to embed that theory in our matching model.

\paragraph{Firms} Our model abstracts from firms and blends the labour and product markets into a single service market. For certain applications, however, it would be useful to introduce firms. This should be straightforward using the framework of \ct{MS13}. In that model, firms hire workers on the labour market and sell goods and services on the product market. The labour and product markets are both organized around a matching function. The extension would introduce a distinction between labour market tightness and product market tightness, between unemployment and idle time, and between wage inflation and price inflation.

\paragraph{Endogenous labour supply} Our model assumes that labour-force participation is fixed. The assumption seems appropriate to describe business cycles \cp[p.~294]{S09}. But to study policies that disincentivize work, such as an income tax, it would be useful to endogenize labour supply. \ct[online appendix D]{MS15} show how to do it.

A related assumption is that the social value of nonwork---the value of recreation and home production while unemployed---is set to 0. The assumption is not unrealistic \cp[pp.~19--21]{MS16}. Nevertheless, it would be interesting to generalize the model to allow for a nonzero value.

\paragraph{Government spending} A common stabilization policy not covered here is government spending. It is easy, however, to introduce government spending into the model \cp{MS15}. Optimal government spending is then given by a formula involving the unemployment gap and government-spending multiplier (the effect of government spending on unemployment). The formula is implementable because estimates of the unemployment gap and government-spending multiplier are available \cp{Ra11,MS16}. The formula shows that optimal government spending reduces but does not eliminate the unemployment gap. In that, government spending differs from monetary policy; the difference arises because government spending distorts households' consumption baskets.

An advantage of the model is that it produces government-spending multipliers that are higher when unemployment is higher \cp{M12,MS15}. It is therefore consistent with the evidence produced by \ct{AG10,AG11}, \ct{CL13}, \ct{FMP12}, \ct{HS11}, and \ct{JT16}. The state-dependence of multipliers generalizes beyond government spending to other fiscal policies, such as the income tax \cp{GZ19}.

\paragraph{Social insurance and redistribution} Social insurance and redistributive policies are also widely used over the business cycle but not covered here. To study these policies, we would need to introduce heterogeneous households into the model. Because the model features unemployment, which is a major source of inequality in modern economies, it is well adapted to such extension. For instance, \ct{LMS15,LMS10} introduce employed and unemployed workers in a variant of the model to study optimal unemployment insurance over the business cycle. \ct{Ko19} adds an heterogeneous-agent structure to study the effects of fiscal policy.

\bibliography{\bib}

\end{document}